\documentclass[preprint,review,12pt]{elsarticle}

\usepackage{graphics}

\usepackage{epsfig}

\usepackage{amssymb}

\usepackage{amsmath}

\usepackage{graphicx}

\usepackage{subfigure}

\usepackage{caption}

\usepackage{booktabs}

\journal{Optics Communications}

\begin{document}

\begin{frontmatter}

\title{Nth-order nonlinear intensity fluctuation amplifier}

\author[a]{Shuanghao Zhang}
\author[a,b]{Huaibin Zheng\corref{Zheng}}
\cortext[Zheng]{Corresponding author.}
\ead{huaibinzheng@xjtu.edu.cn}
\author[c]{Gao Wang}
\author[a]{Hui Chen}
\author[a]{Jianbin Liu}
\author[d]{Yu Zhou}
\author[a]{Yuchen He}
\author[a]{Sheng Luo}
\author[b]{Yanyan Liu}
\author[a]{Zhuo Xu}

\address[a]{Key Laboratory of Multifunctional Materials and Structures, Ministry of Education, School of Electronic Science and Engineering, Xi'an Jiaotong University, Xi'an, Shaanxi 710049, China}

\address[b]{Science and Technology on Electro-Optical Information Security Control Laboratory, Tianjin 300308, China}

\address[c]{School of Physics $\&$ Astronomy, University of Glasgow, Glasgow G12 8QQ, UK}

\address[d]{MOE Key Laboratory for Nonequilibrium Synthesis and Modulation of Condensed Matter, Department of Applied Physics, Xi'an Jiaotong University, Xi'an 710049, China}

\begin{abstract}
Stronger light intensity fluctuations are pursued by related applications such as optical resolution, image enhancement, and beam positioning. In this paper, an Nth-order light intensity fluctuation amplifier is proposed, which was demonstrated by a four-wave mixing process with different statistical distribution coupling lights. Firstly, its amplification mechanism is revealed both theoretically and experimentally. The ratio $R$ of statistical distributions and the degree of second-order coherence ${g^{(2)}}(0)$ of beams are used to characterize the affected modulations and the increased light intensity fluctuations through the four-wave mixing process. The results show that the amplification of light intensity fluctuations is caused by not only the fluctuating light fields of incident coupling beams, but also the fluctuating nonlinear coefficient of interaction. At last, we highlight the potentiality of applying such amplifier to other N-order nonlinear optical effects. 
\end{abstract}

\end{frontmatter}

\section{Introduction}
\label{sec:Introduction}

Controlling light intensity fluctuations has been demonstrated to be a topic of pivotal importance in many optical applications. In the process of imaging, the useful information can be contained not only in the light field but also in the light intensity fluctuations \cite{kolobov2007quantum,zeng2013laser,boyer2008entangled}. The role of light intensity fluctuations in improving the resolution of optical image has been proved \cite{PhysRevLett.85.3789,sprigg2016super,sroda2020sofism,DIGMAN2005L33}. The impact of light intensity fluctuations on the visibility of ghost imaging, ghost interference, and ghost diffraction were reported in \cite{ferri2005high,liu2013high,bromberg2009ghost,shapiro2008computational,gatti2004ghost}. The beam positioning was performed by the fluctuation of an actual beam oscillation \cite{treps2003quantum}. The fluctuations of the light field were widely utilized to identify diverse sources of light \cite{PhysRev.130.2529,mandel1995optical,you2020identification}. The optical rogue waves and extreme events are defined as an optical pulse whose amplitude or intensity is much higher than that of the surrounding pulses, which were closely related to the light intensity fluctuations \cite{PhysRevLett.119.223603,akhmediev2016roadmap}. The reference \cite{zhang2019superbunching} showed that the bunching property of the light field can be modified via modulation of the intensity fluctuation. In the presence of the fluctuating light pump, the multiphoton absorption \cite{lambropoulos1966coherence,jechow2013enhanced}, multiphoton ionization \cite{lecompte1975laser,mouloudakis2019revisiting}, generation of optical harmonics \cite{PhysRevLett.119.223603,lamprou2020perspective}, and modulation instability \cite{Hammani:09} were hugely enhanced. Therefore, stronger light intensity fluctuation plays an important role in many optical applications.

In general, the amplification of the light intensity fluctuations has been achieved in both linear and nonlinear methods. In the linear optical system, the increased intensity fluctuations were obtained by rotating ground glasses \cite{zhou2017superbunching} and adding electro- or acousto-optical modulator \cite{straka2018generator,zhou2019superbunching}. However, due to experimental complexity and error resulted from the increased number of linear devices, the amplification degree of the intensity fluctuations is limited in linear methods.

In the nonlinear optical system, the light intensity fluctuations were enhanced in the interaction of matters and light \cite{perina1993photon,cao2016resolution,Yu:16}, squeezing processes \cite{liu2016enhanced}, and rogue waves or extreme events \cite{PhysRevLett.119.223603,manceau2019indefinite}. However, the amplification degree of light intensity fluctuations in nonlinear processes is disproportionate and heavily dependent on particular matters and environmental factors that triggered nonlinear interaction.

In this paper, an amplification mechanism for light intensity fluctuations is investigated in an N-order nonlinear optical process both theoretically and experimentally. The Nth-order light intensity fluctuation amplifier is achieved by incident beams configures with different statistical distributions involving in an N-order nonlinear optical effects. Four-wave mixing (FWM) was chosen as verification of this amplifier, 1$\sim$N incident beams were modulated simultaneously into the pseudothermal light through the FWM process, which introduced and compared the Nth-order light intensity fluctuation amplifier. The ratio $R$ of statistical distributions and the degree of second-order coherence ${g^{(2)}}(0)$ of beams through FWM were used to evaluate the affected modulations and the increased light intensity fluctuations.

Inputting light beams configures with different statistical distributions into N-order optical nonlinear processes brings different modulations and light intensity fluctuations. In the proposed Nth-order amplifier, the amplification degree of light intensity fluctuations can be tunable by different statistical distributions configures, and the multiplex amplification can be simultaneously achieved. On the other hand, by different statistical distributions involving in the N-order nonlinear optical process, the nonlinearity changes accordingly. Thus, the amplification of light intensity fluctuations is caused by not only the fluctuating light fields of incident coupling beams, but also the fluctuating nonlinear coefficient of interaction. In principle, the amplifier can be widely used in many N-order nonlinear optical effects, including optical harmonics, electromagnetically induced transparency, and four-wave mixing. The proposal highlights the potential applications as a flexible and powerful method to amplify the intensity fluctuation of light. 

\section{Theory}
\label{sec:Theory}

In the N-order nonlinear processes, the optical field in atomic systems can be described in terms of equations of motion \cite{boyd2003nonlinear}
\begin{eqnarray}
\tilde {E}(t) =  {E}{e^{ - i\omega t}} + c.c.,\
\label{eq:one}
\end{eqnarray}
where ${E}$ denotes the complex amplitude of the arbitrary wave at frequency $\omega$. The nonlinear interaction response can often be described by expressing the polarization $\tilde {P}(t)$ as
\begin{eqnarray}
\tilde P(t) = {\varepsilon _0}\left[ {{\chi ^{(1)}}\tilde E(t) + {\chi ^{(2)}}{{\tilde E}^2}(t) +  \ldots  + {\chi ^{(N)}}{{\tilde E}^N}(t)} \right].\
\label{eq:two}
\end{eqnarray}
${\chi ^{\left( 1 \right)}}\left( \omega  \right)$ is linear susceptibility and ${\chi ^{\left( N \right)}}\left( \omega  \right)$ is N-order nonlinear susceptibility. As seen that, there are N incident light fields $\tilde {E}(t)$ involved in the N-order nonlinear process, and the nonlinearity results from the interaction of matters and N incident beams. Thus, the output light through N-order nonlinear processes will exhibit different modulation results affected by the nonlinearity from different incident light fields combinations.

The third-harmonic generation and FWM are the third-order nonlinear processes, we use a degenerate FWM process to explain and introduce the proposed Nth-order light intensity amplifier. In the FWM process, we use different incident light fields combinations to manipulate nonlinear interaction results. The ratio $R$ of the statistical light intensity distribution and the degree of second-order coherence ${g^{(2)}}(0)$ of beams through FWM were used to evaluate the affected modulations and the increased light intensity fluctuations. The ratio $R$ is defined as
\begin{eqnarray}
R = \frac{{{{(\Delta I)}^2}}}{{{{\bar I}^2}}},\
\label{eq:three}
\end{eqnarray}
where $(\Delta I{)^2}$ and $\bar I$ are the variance and mean value, respectively. For the degree of the second-order coherence, the normalized second-order coherence functions of beams involved in the FWM process can be calculated as \cite{mandel1995optical}
\begin{eqnarray}
{g^{(2)}}(\tau ) = \frac{{\left\langle {\bar I(t)\bar I(t + \tau )} \right\rangle }}{{{{\left\langle {\bar I(t)} \right\rangle }^2}}},\
\label{eq:four}
\end{eqnarray}
where $\tau$ denotes the delay time, $\bar I(t)$ is the average of the intensity over a cycle of oscillation and the brackets $\left\langle  \ldots  \right\rangle$ denote the statistical or longer-time average. The ${g^{(2)}}(\tau)$ when $\tau=0$ is defined as the degree of second-order coherence ${g^{(2)}}(0)$ \cite{Loudon1983The}. 

In the degenerate FWM process, the wave vector mismatch is zero in the presence of the phase matching condition. The amplitude ${E_i}$ of any field propagating in the $+z$ direction obeys the set of coupled equations
\begin{eqnarray}
\frac{{d{E_i}}}{{dz}} =  - \alpha {E_i} + \kappa E_j^*.\
\label{eq:five}
\end{eqnarray}
${E_j}$ is the amplitude of phase conjugation wave of ${E_i}$. The absorption coefficient $\alpha$ and coupling coefficient $\kappa$ are 
\begin{subequations}
\label{eq:six}
\begin{equation}
\alpha  =  - \frac{\omega }{{2nc}}{\mathop{\rm Im}\nolimits} {\chi ^{(1)}}(\omega ),\label{subeq:1}
\end{equation}
\begin{eqnarray}
\kappa  =  - i\frac{{3\omega }}{{2nc}}{\chi ^{(3)}}(\omega ){E_m}{E_n}.\label{subeq:2}
\end{eqnarray}
\end{subequations}
where $n = \sqrt {1 + {\mathop{\rm Re}\nolimits} {\chi ^{(1)}}(\omega )}$ is the usual linear refractive index and $c = 3 \times {10^8}m/s$ is the speed of light in vacuum. ${E_m}$ and ${E_n}$ are the amplitudes of two pump waves. The linear susceptibility ${\chi ^{\left( 1 \right)}}\left( \omega  \right)$ and the third-order nonlinear susceptibility ${\chi ^{\left( 3 \right)}}\left( \omega  \right)$ are given by
\begin{subequations}
\label{eq:seven}
\begin{equation}
{\chi ^{\left( 1 \right)}}\left( \omega  \right) = \frac{{N{\mu ^{\rm{2}}}\omega _{\rm{0}} \hbar }}{{{\varepsilon _{\rm{0}}}}}\frac{{{T_2}\left( { - i + \Delta {T_2}} \right)}}{{{\hbar ^2}\left( {1 + {\Delta ^2}T_2^2} \right) + 4{T_1}{T_2}{\mu ^{\rm{2}}}{E_m}{E_n}}},\label{subeq:3}
\end{equation}
\begin{eqnarray}
{\chi ^{\left( 3 \right)}}\left( \omega  \right) =  - \frac{{4N{\mu ^4}\omega _{\rm{0}} }}{{3{\varepsilon _{\rm{0}}}\hbar \left( {i + \Delta {T_2}} \right)}}\frac{{{T_1}T_2^2}}{{{\hbar ^2}\left( {1 + {\Delta ^2}T_2^2} \right) + 4{T_1}{T_2}{\mu ^{\rm{2}}}{E_m}{E_n}}}.\label{subeq:4}
\end{eqnarray}
\end{subequations}

In this experiment, due to definite nonlinear medium and atom level of the D2 line $(5{S_{1/2}}(F = 2) \leftrightarrow 5{P_{3/2}}(F' = 3))$ of ${}^{{\rm{87}}}Rb$ \cite{boyd2003nonlinear,steck2001rubidium}, all atomic parameters become invariant constants. The transition dipole moment $\mu {\rm{ = 1}}{\rm{.731}} \times {\rm{1}}{{\rm{0}}^{{\rm{ - 29}}}}Cm$, longitudinal and transverse relaxation time are ${T_1} = 2.62357 \times 10{}^{ - 8}s$ and ${T_2} = 5.24714 \times 10{}^{ - 8}s$. The Planck's constant and permeability of vacuum is $\hbar  = 1.054 \times {10^{ - 34}}Js$ and ${\varepsilon _0} = 8.854 \times {10^{ - 12}}F/m$, respectively. For the degenerate FWM, the detuning $\Delta$ of the laser frequency from the resonant frequency equals $0$. The laser angular frequency is $\omega  = {\omega _0} = 2\pi  \cdot 384.230THz$. The atomic density of Rb vapor is approximately $N = 6.3 \times {10^{12}}c{m^{ - 3}}$ (cell temperature $95{}^ \circ C$) \cite{steck2001rubidium}. By combining the above constant parameters and Eq.~(\ref{eq:seven}), we reduce Eq.~(\ref{eq:six}) to
\begin{subequations}
\label{eq:eight}
\begin{equation}
\alpha  = \frac{{1.14 \times {{10}^{ - 47}}}}{{1.11 \times {{10}^{ - 68}} + 1.65 \times {{10}^{ - 72}}{E_m}{E_n}}},\label{subeq:5}
\end{equation}
\begin{eqnarray}
\kappa  = \frac{{1.7 \times {{10}^{ - 51}}{E_m}{E_n}}}{{1.11 \times {{10}^{ - 68}} + 1.65 \times {{10}^{ - 72}}{E_m}{E_n}}}.\label{subeq:6}
\end{eqnarray}
\end{subequations}

As seen in Eq.~(\ref{eq:eight}), if the atomic and environmental conditions of the nonlinear process are defined, the nonlinear coefficients will be dependent on the light fields of pump beams ${E_m}$ and ${E_n}$. Thus the affected modulation of the light field ${E_i}$ after nonlinear interaction is the result of combining its phase conjugation field ${E_j}$ with two pump light fields ${E_m}$ and ${E_n}$. Here the probe and pump light fields in the FWM process are considered as the light fields with different fluctuations. The output beam through nonlinear interaction is modulated by not only the fluctuating light fields of incident coupling beams but also fluctuating nonlinear coefficient. Based on the above analysis, the affected modulation and light intensity fluctuations of the output beam would be considerable and controllable. 

The amplitude ${E_i}$ of any field at the exit plane of the nonlinear medium is obtained by integral operation, then the intensity ${I_i}$ of any beam involved is given by 
\begin{eqnarray}
{I_i} = 2n{\varepsilon _0}c{\left| {{E_i}} \right|^2}.\
\label{eq:nine}
\end{eqnarray}
Then the ratio $R$ of the statistical light intensity distribution and the degree of second-order coherence ${g^{(2)}}(0)$ of beams through FWM were measured to evaluate the affected modulations and the increased light intensity fluctuations.

\section{Experimental setup}
\label{Experimental setup}

\begin{figure}[htbp]
\centering
\includegraphics[width=0.85\textwidth]{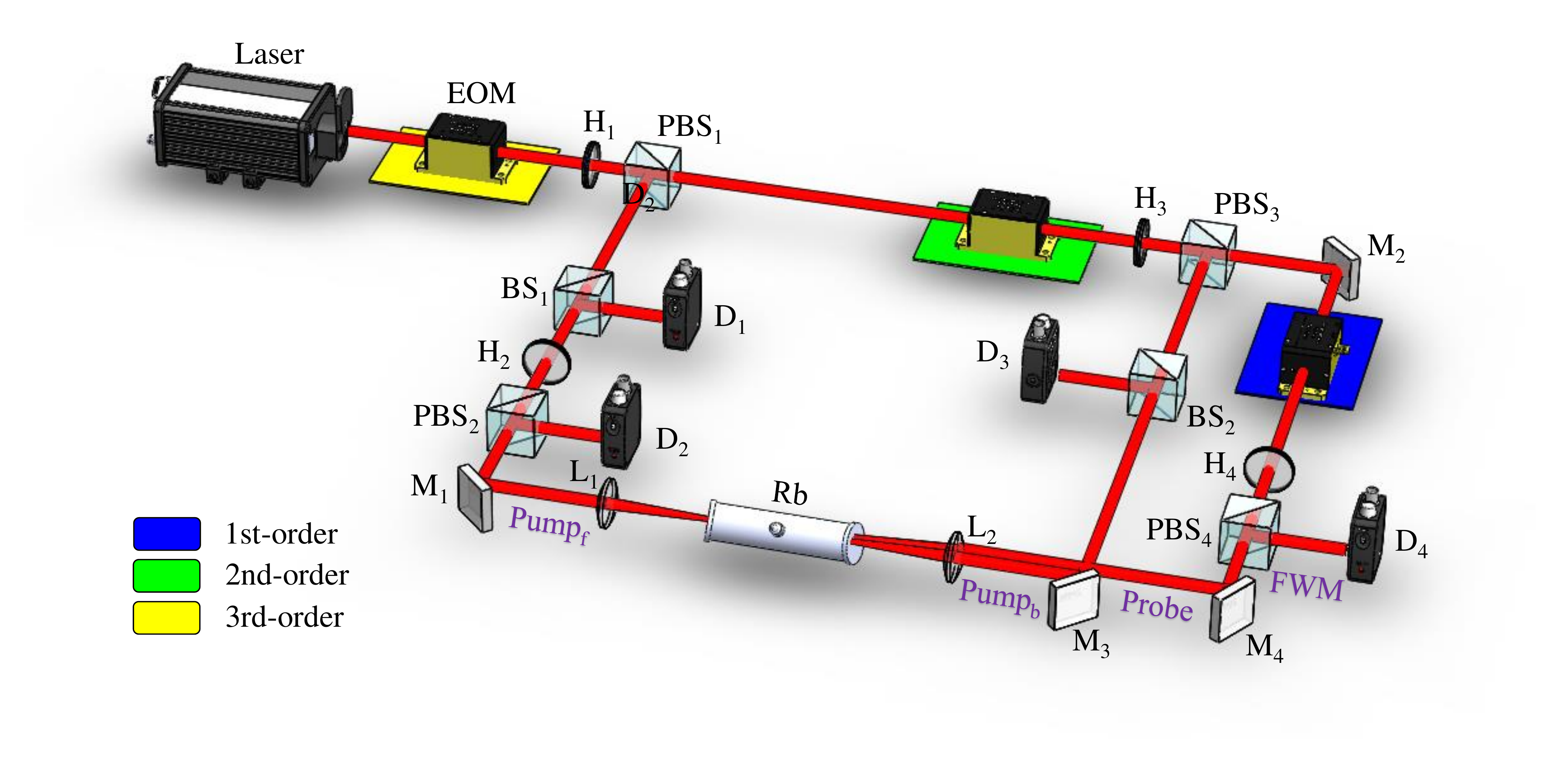}
\caption{\label{fig:1} Schematics of the experimental setup. Placing an EOM in different color area corresponds to the Nth-order intensity fluctuations amplifier. Laser: single-mode tunable diode laser; EOM: electro-optical modulator; L$_1$-L$_2$: convex lenses; M$_1$-M$_4$: mirrors; PBS$_1$-PBS$_4$: polarization beam splitters; H$_1$-H$_4$: half-wave plates; D$_1$-D$_4$: high-speed photoelectric detectors; BS$_1$-BS$_2$: beam splitters; Rb: rubidium vapor cell.}
\end{figure}
The experimental setup of the Nth-order light intensity fluctuation amplifier is schematically depicted in Fig.~\ref{fig:1}. The laser was from a single-mode tunable diode laser with a center wavelength of $780.24 nm$ and power of $56.28 mW$. The laser beam was divided into two beams by a set of half-wave plate (H$_1$) and polarization beam splitter (PBS$_1$). The reflected beam with S-polarization component was calibrated into P-polarization component via the second set of half-wave plate (H$_2$) and polarization beam splitter (PBS$_2$), named as forward pump beam (Pump$_f$). Meantime, the transmitted beam was further divided into two parts by a set of half-wave plate (H$_3$) and polarization beam splitter (PBS$_3$). The reflected part with S-polarization served as the backward pump beam (Pump$_b$) and counter-propagated with the Pump$_f$. The transmitted part through half-wave plate (H$_4$) and polarization beam splitter (PBS$_4$) was chosen as the probe beam with P-polarization. The small angle between probe beam and the backward pump beam is ${0.7^ \circ }$. The two convex lenses (L$_1$ and L$_2$) were used to focus three incident beams at a point in the rubidium (Rb) atomic vapor cell. Since they satisfied the phase-matching condition of a degenerated-FWM process, the new FWM signal was generated from the opposite direction of the probe beam with S-polarization. 

The Rb cell was set to $95^\circ C$ by a temperature control heater, which had a length of $75 mm$. The forward and backward pump beam power were $12.06 mW$ and $5.24 mW$, the probe power was $8.14 mW$. After the FWM process, the high-speed photoelectric detectors (D$_1$ and D$_3$) were employed to detect the light intensities of the Probe and Pump$_f$ via the beam splitters (BS$_1$ and BS$_2$), respectively. The light intensities of the Pump$_b$ and the FWM signal were measured by the D$_2$ and D$_4$.

In this experiment, the pseudothermal light was obtained through an electro-optical modulator (EOM) and by inputting exponential distribution signal. As shown in Fig.~\ref{fig:1}, by placing an EOM in different-colors locations, one or more incident beams were modulated simultaneously into the pseudothermal light through the FWM process, that realized the Nth-order light intensity fluctuation amplifier. In our schemes, we used three cases to introduce and compare the Nth-order light intensity fluctuation amplifier, the ratio of the statistical light intensity distribution $R$ (accurate to two decimal places) and the degree of second-order coherence ${g^{(2)}}(0)$ of beams through FWM were used to evaluate the affected modulations and the increased light intensity fluctuations.

\section{Results and discussion}
\label{Results and discussion}

\begin{figure}[htbp]
\centering
\includegraphics[width=0.75\textwidth]{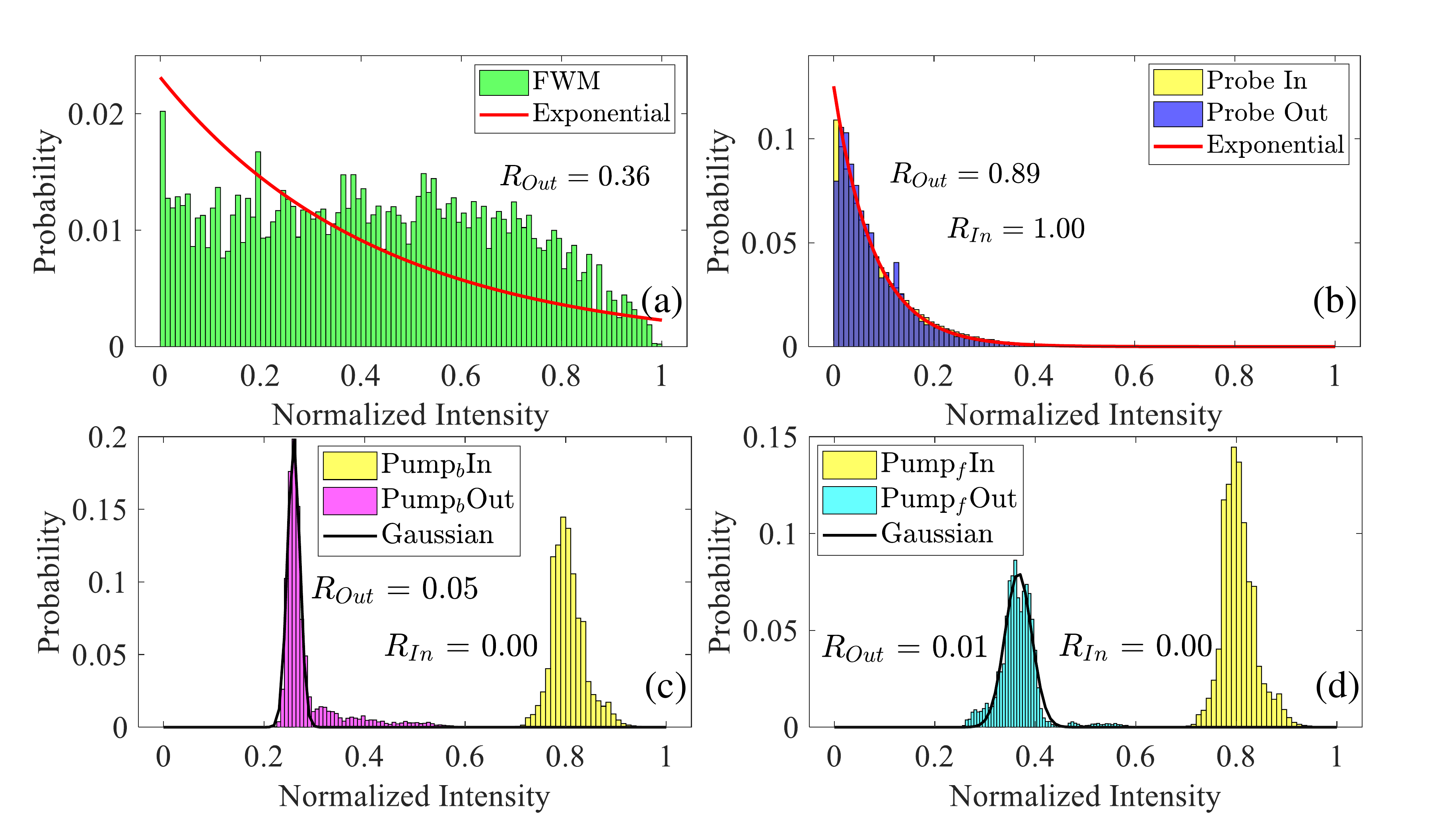}
\caption{\label{fig:2} The measured statistical intensity distributions of (a) FWM signal, (b) probe beam, (c) backward and (d) forward pump beams through FWM in the 1st-order case. The red and black lines represent the exponential and Gaussian distributions with the same average intensities of output beam, respectively.}
\end{figure}

In the 1st-order case, an EOM is placed in the blue area as shown in Fig.~\ref{fig:1}. The probe beam is modulated to thermal light with $R_{In}=1.00$ by EOM, the forward and backward pump beams are lasers with $R_{In}=0.00$. As seen that, two incident beams through FWM process are lasers in the 1st-order case. Figure~\ref{fig:2} shows the statistical intensity distributions of beams before and after the FWM process in the 1st-order case. The light intensity distribution of the generated FWM signal is shown in Fig.~\ref{fig:2}(a). Compared with the exponential distribution with the same average intensity (red line), the FWM signal shows a heavy-tailed distribution with $R_{Out}=0.36$ according the Ref. \cite{manceau2019indefinite}. As shown in Fig.~\ref{fig:2}(b), the original probe beam is the thermal light obeying negative exponential distribution with $R_{In}=1.00$. There is very little modulation in the output of the probe beam after the FWM process, whose statistical intensity distribution remain largely the exponential distribution with $R_{Out}=0.89$. As shown in Figs.~\ref{fig:2}(c) and \ref{fig:2}(d), the incident forward and backward pump beams are laser, whose statistical intensity distributions are Gaussian distributions with $R_{In}=0.00$. Even the intensities of backward and forward pump beams decreased after the FWM process, their statistical intensity distributions still exhibit profiles of Gaussian distributions with $R_{Out}=0.05$ and $R_{Out}=0.01$. The black lines represent the Gaussian distributions with the same average intensities. It revealed that the configurations of incident light fields in the 1st-order case had little effect on amplifying affected modulation and light intensity fluctuation. 
\begin{figure}[t]
\centering
\includegraphics[width=0.75\textwidth]{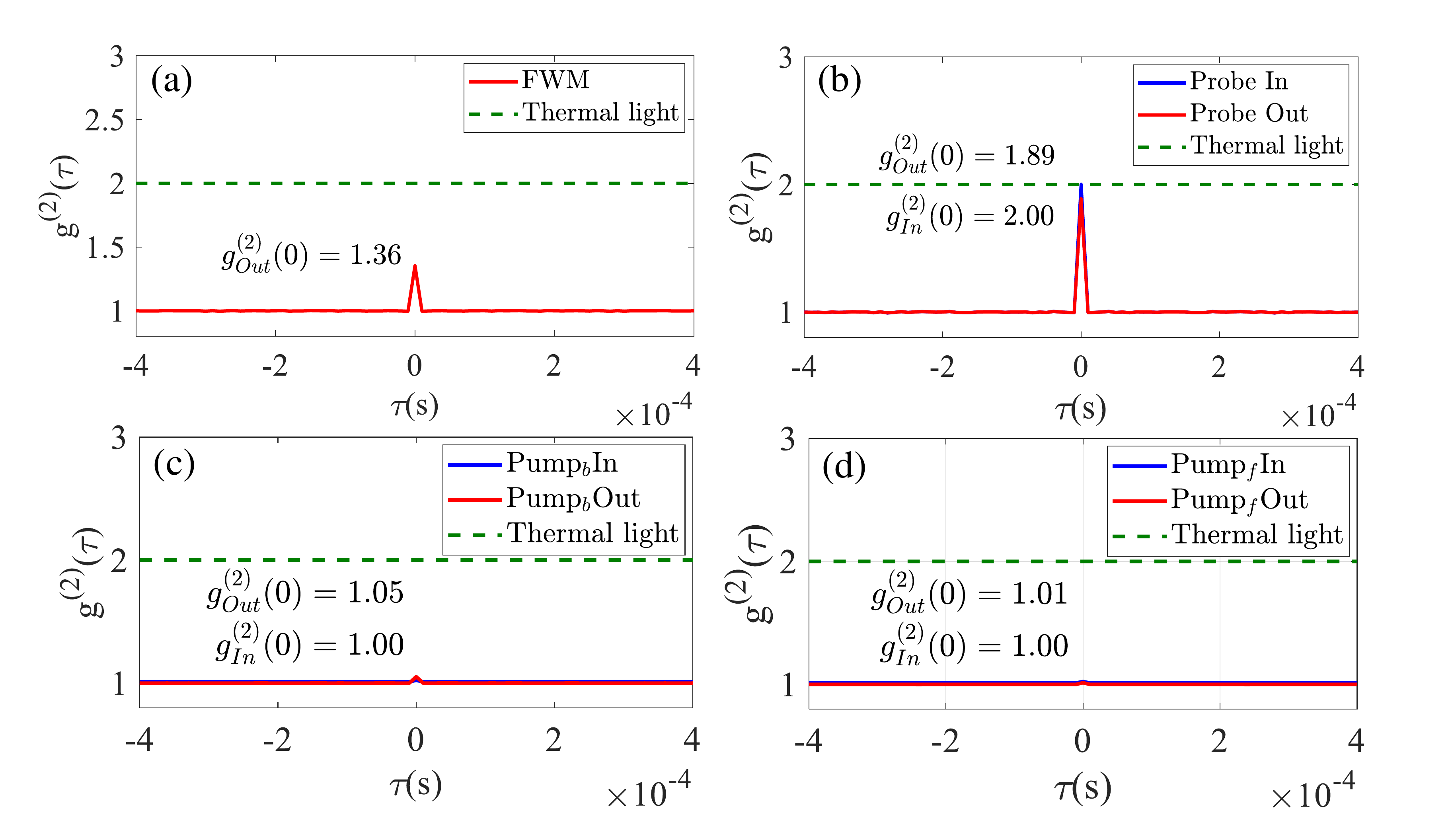}
\caption{\label{fig:3} The measured second-order coherence functions of (a) FWM signal, (b) probe beam, (c) backward and (d) forward pump beams in the 1st-order case. Dashed green lines represent the normalized degree of second-order coherence ${g^{(2)}}(0)$ of thermal light in theory, that is boundary between bunching and superbunching effect. Solid blue lines are the measured second-order coherence functions of input beams experimentally. Solid red lines indicate the measured second-order coherence functions of output beams through FWM process.}
\end{figure}

Figure~\ref{fig:3} shows the measured second-order coherence functions of beams through the FWM process in the 1st-order case. The degree of second-order coherence ${g^{(2)}}(0)$ of the generated FWM signal reached $1.36$ in Fig.~\ref{fig:3}(a). The ${g^{(2)}}(0)$ of probe beam decreased lightly from $2.00$ to $1.89$ in Fig.~\ref{fig:3}(b). The backward and forward pump beams through FWM process basically maintain ${g^{(2)}}(0)=1.00$ in Figs.~\ref{fig:3}(c) and \ref{fig:3}(d). As seen that, the configurations of incident light fields in the 1st-order case were ineffective in improving the degree of second-order coherence of beams.
\begin{figure}[t]
\centering
\includegraphics[width=0.75\textwidth]{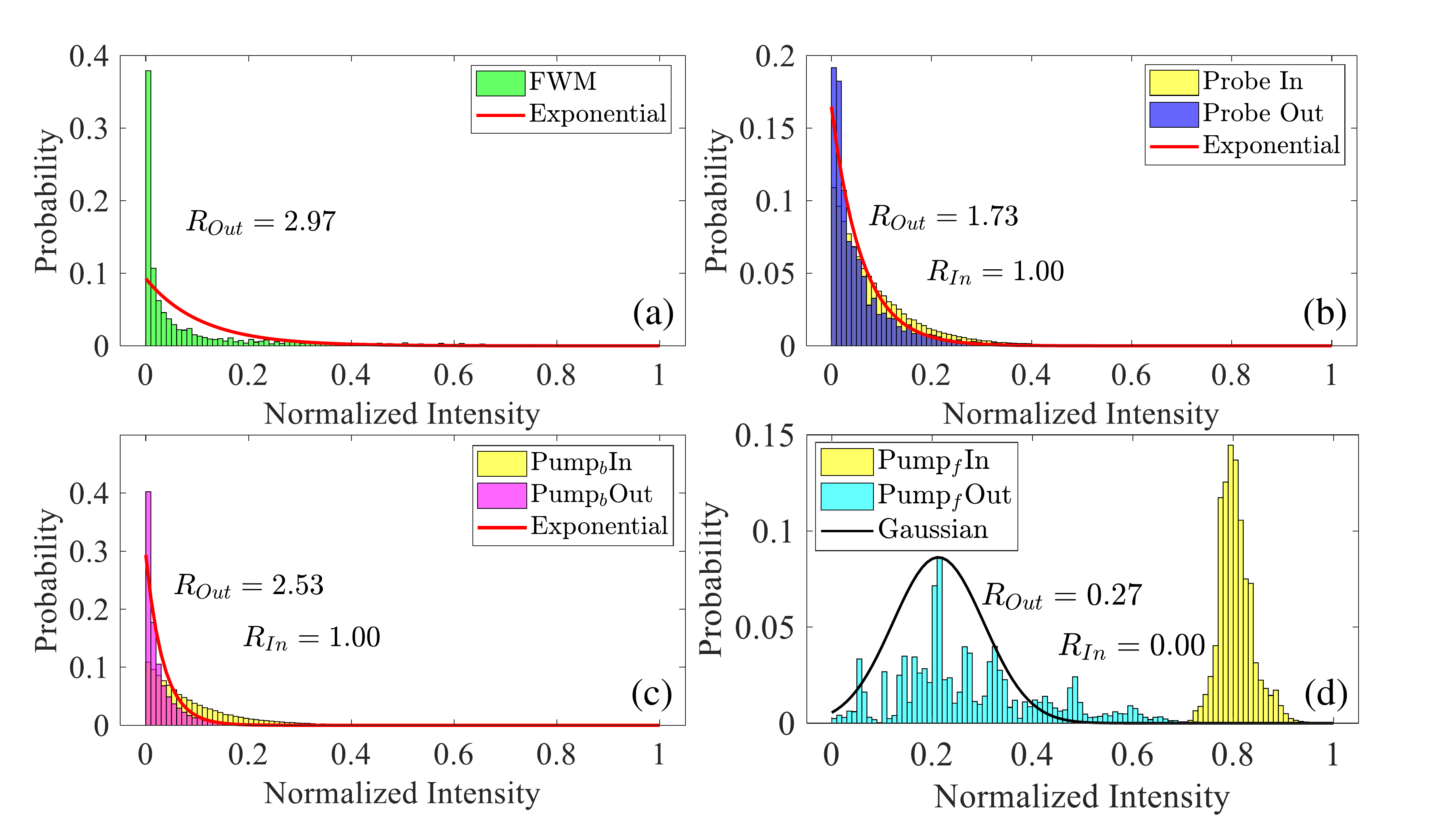}
\caption{\label{fig:4} The measured statistical intensity distributions of (a) FWM signal, (b) probe beam, (c) backward and (d) forward pump beams through FWM in the 2nd-order case. The red and black lines represent the exponential and Gaussian distributions with the same average intensities of output beam, respectively.}
\end{figure}

In the 2nd-order case, an EOM is placed in the green area as shown in Fig.~\ref{fig:1}. The probe beam and backward pump beam were modulated simultaneously to the same thermal light with $R_{In}=1.00$ by EOM, the forward pump is laser with $R_{In}=0.00$. As seen that, only one incident beam through FWM process is laser in the 2nd-order case. Figure~\ref{fig:4} shows the statistical intensity distributions of beams before and after the FWM process in the 2nd-order case. The light intensity distribution of the generated FWM signal is shown in Fig.~\ref{fig:4}(a). Compared with the exponential distribution with the same average intensity (red line), the FWM signal shows heavy-modulated statistical distribution and has strong fluctuation with $R_{Out}=2.97$. As shown in Fig.~\ref{fig:4}(b), the original probe beam is the thermal light obeying negative exponential distribution with $R_{In}=1.00$. After the FWM process, the statistical intensity distribution of the probe beam is modulated slightly with $R_{Out}=1.73$. As shown in Fig.~\ref{fig:4}(c), the forward pump beam is changed from the exponential distribution with $R_{In}=1.00$ to the heavy-modulated distribution with $R_{Out}=2.53$. The incident forward pump beam is laser obeying Gaussian distribution with $R_{In}=0.00$, yet the output basically maintains Gaussian-like distribution with $R_{Out}=0.27$ as shown in Fig.~\ref{fig:4}(d). It is concluded that the statistical intensity distributions of beams in the 2nd-order case are deeper modulated than the 1st-order case, and the light intensity fluctuations of all beams are amplified.
\begin{figure}[htbp]
\centering
\includegraphics[width=0.75\textwidth]{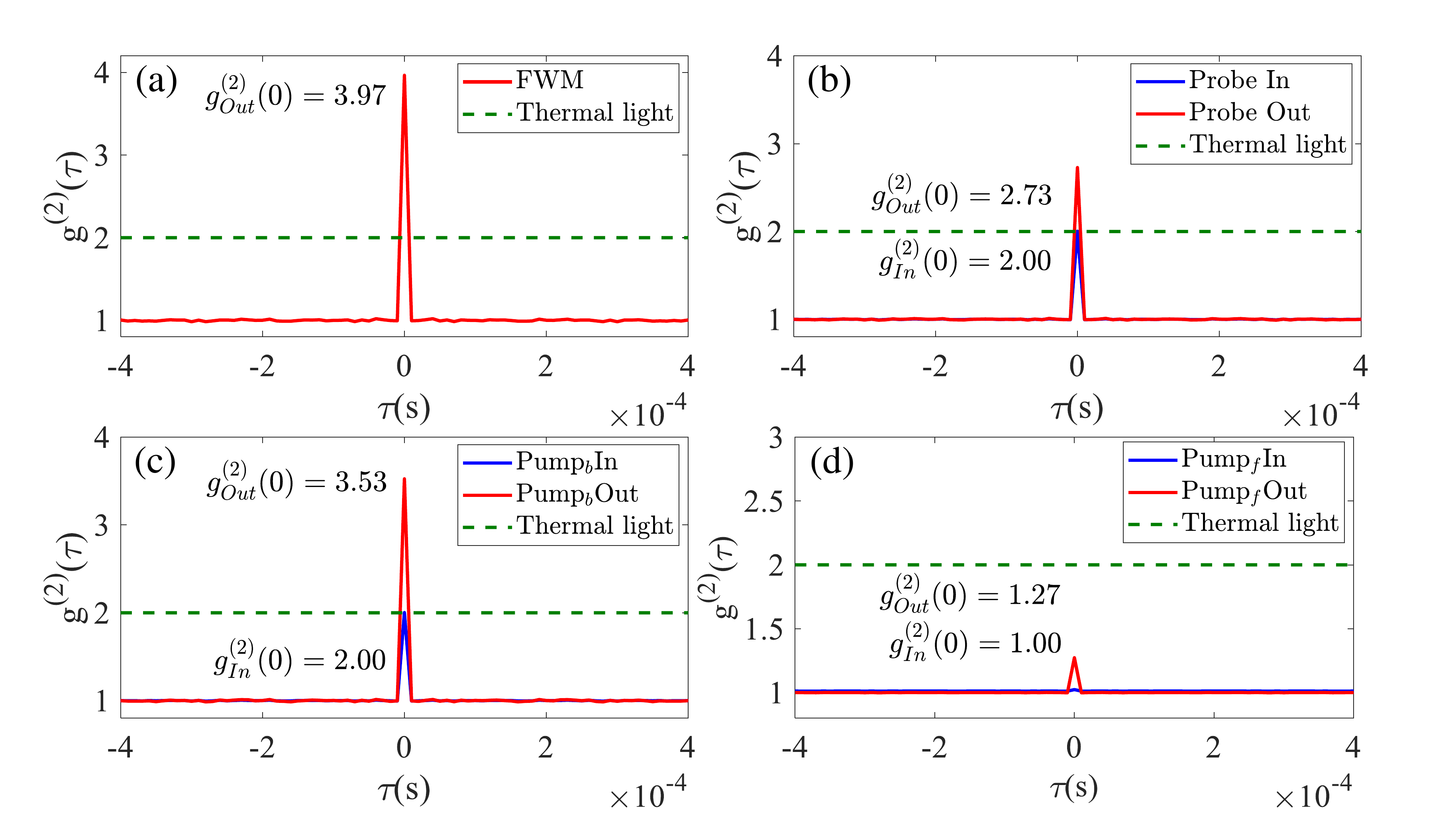}
\caption{\label{fig:5} The measured second-order coherence functions of (a) FWM signal, (b) probe beam, (c) backward and (d) forward pump beams in the 2nd-order case. Dashed green lines represent the normalized degree of second-order coherence ${g^{(2)}}(0)$ of thermal light in theory, that is the boundary between bunching and superbunching effect. Solid blue lines are the measured second-order coherence functions of input beams experimentally. Solid red lines indicate the measured second-order coherence functions of output beams through the FWM process.}
\end{figure}

Figure~\ref{fig:5} shows the second-order coherence functions of beams through the FWM process in the 2nd-order case. The FWM signal with ${g^{(2)}}(0)=3.97$ exceeds the superbunching threshold value $2.00$ in Fig.~\ref{fig:5}(a). The ${g^{(2)}}(0)$ of probe beam slightly increased from $2.00$ to $2.73$ in Fig.~\ref{fig:5}(b). The ${g^{(2)}}(0)$ of backward pump beam substantially increased from $2.00$ to $3.53$ in Fig.~\ref{fig:5}(c). The reason for different growth of ${g^{(2)}}(0)$ is that the beam with strong power is hardly modulated deeply. The forward pump beams slightly ascended to ${g^{(2)}}(0)=1.27$ from $1.00$ in Fig.~\ref{fig:5}(d). As seen that, the FWM signal, probe beam, and backward beam through the FWM process all reached superbunching effects in the 2nd-order case \cite{boitier2011photon,iskhakov2012superbunched}. We derive a conclusion that the incident light fields configurations of the 2nd-order case had done better than the 1st-order case in improving the degree of second-order coherence of beams.
\begin{figure}[t]
\centering
\includegraphics[width=0.75\textwidth]{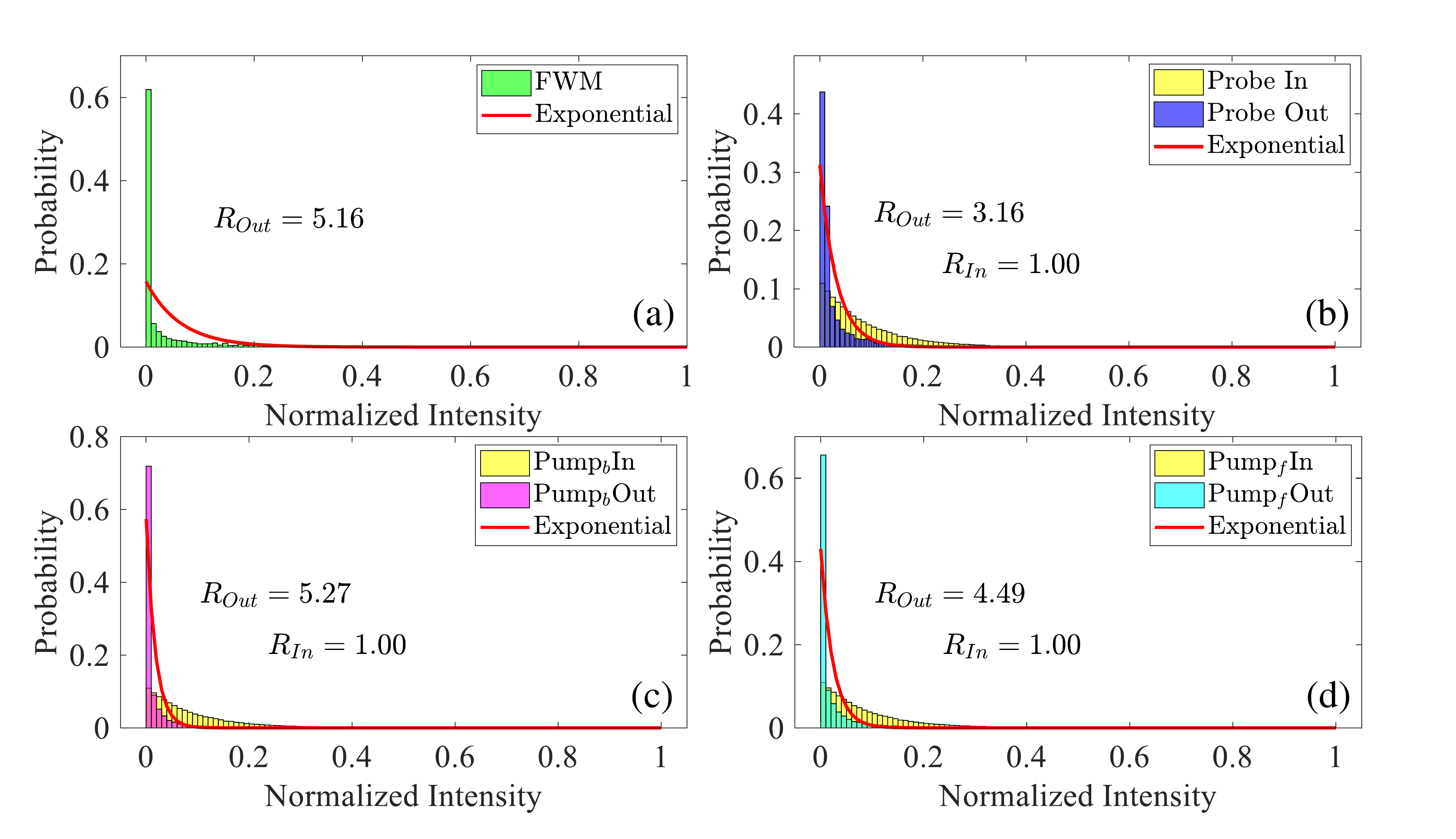}
\caption{\label{fig:6} The measured statistical intensity distributions of (a) FWM signal, (b) probe beam, (c) backward and (d) forward pump beams through FWM in the 3rd-order case. Red lines represent the exponential distribution with the same average intensity of output beam.}
\end{figure}

In the 3rd-order case, an EOM is placed in the yellow area as shown in Fig.~\ref{fig:1}. All incident beam are simultaneously modulated to the same thermal light by EOM. As seen that, there is no laser through FWM process in the 3rd-order case. Figure~\ref{fig:6} shows the statistical intensity distributions of involves beams before and after the FWM process in the 3rd-order case. The light intensity distribution of the generated FWM signal is shown in Fig.~\ref{fig:6}(a). Compared with exponential distribution with the same average intensity (red line), the FWM signal shows heavy-modulated distribution with $R_{Out}=5.16$. As shown in Fig.~\ref{fig:6}(b), \ref{fig:6}(c) and \ref{fig:6}(d), the probe, backward and forward pump beam is changed from the exponential distribution with $R_{In}=1.00$ to the heavy-modulated distribution with $R_{Out}=3.16,5.27,4.49$, respectively. It is demonstrated that the simultaneously fluctuating light configuration in the 3rd-order case brings the greatest modulation and the strongest intensity fluctuation amplification.
\begin{figure}[!htbp]
\centering
\includegraphics[width=0.75\textwidth]{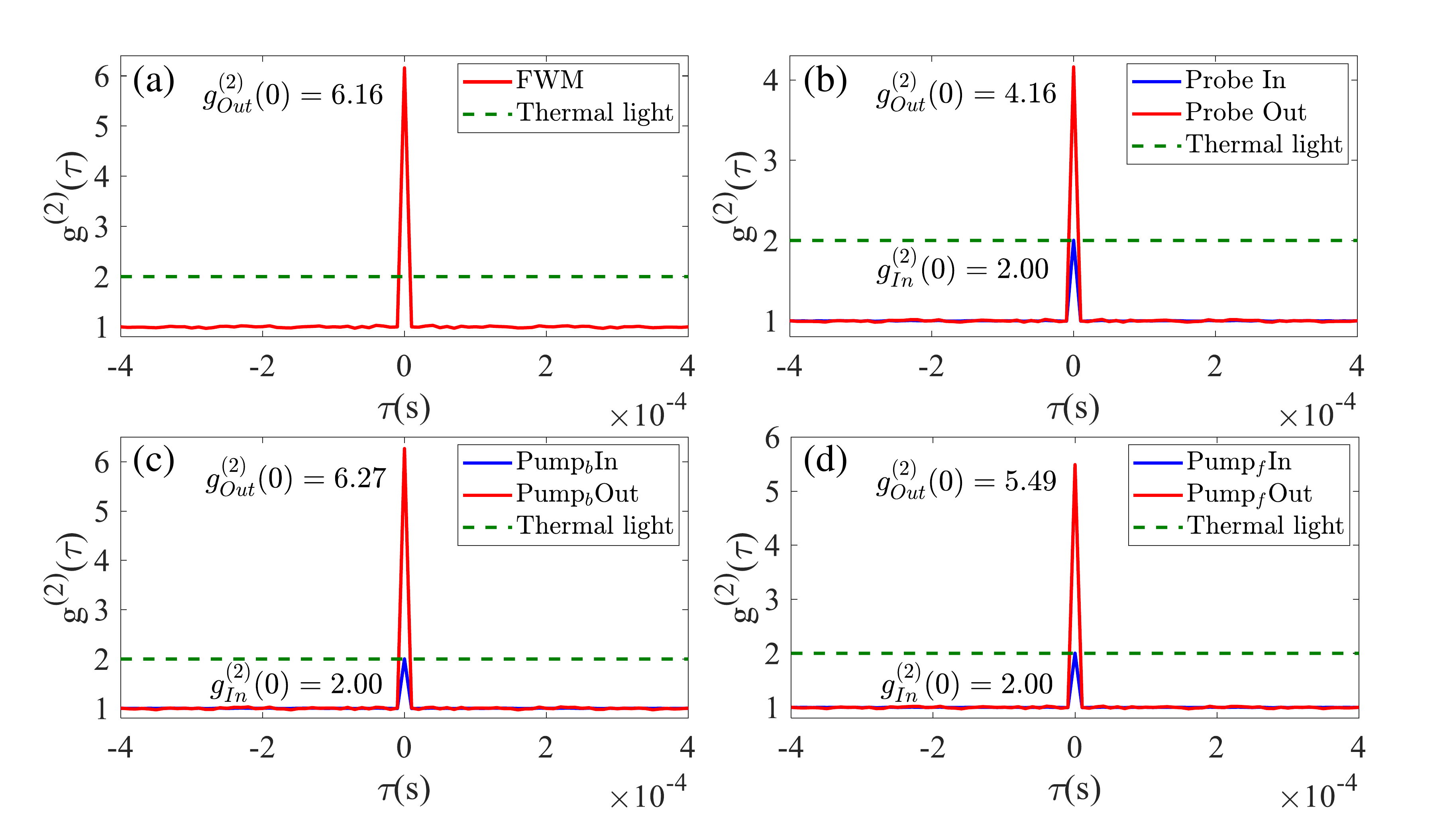}
\caption{\label{fig:7} The measured second-order coherence functions of (a) FWM signal, (b) probe beam, (c) backward and (d) forward pump beams in the 3rd-order case. Dashed green lines represent the normalized degree of second-order coherence ${g^{(2)}}(0)$ of thermal light in theory, that is the boundary between bunching and superbunching effect. Solid blue lines are the measured second-order coherence functions of input beams experimentally. Solid red lines indicate the measured second-order coherence functions of output beams through FWM process.}
\end{figure}

The second-order coherence functions of beams in the 3rd-order case are as shown in Fig.~\ref{fig:7}. The FWM signal, probe, backward, and forward pump beams with ${g^{(2)}}(0)=6.16, 4.16, 6.27, 5.49$ all reached the superbunching effect. The degree of second-order coherence of beams in the 1$\sim$N order cases are summarized in Table~\ref{tab:table2}. As seen that, the configurations of incident light fields in the 2nd-order and the 3rd-order cases achieve great performances in amplifying light intensity fluctuation and improving the degree of second-order coherence. In conclusion, in the third-order ($N$) nonlinear process, zero and one ($k$, $0 \le k < 2/N$) incident beams are the laser or coherent light (the 2nd-order and 3rd-order case), the light intensity fluctuations of all output beams after the nonlinear interaction are enhanced.
\begin{table}[t]
\centering
\caption{\label{tab:table2} The degree of second-order coherence ${g^{(2)}}(0)$ of beams through FWM in the 1$\sim$N order cases.}
\begin{tabular}{ccccc}
\toprule
${g^{(2)}}(0)$ &\mbox{$FWM$} &\mbox{$Probe$} &\mbox{$Pum{p_b}$} &\mbox{$Pum{p_f}$}\\
\midrule
$Case 1$ & 1.36 &\mbox{1.89} &\mbox{1.05} &\mbox{1.01}\\
$Case 2$ & 3.97 &\mbox{2.73} &\mbox{3.53} &\mbox{1.27}\\
$Case 3$ & 6.16 &\mbox{4.16} &\mbox{6.27} &\mbox{5.49}\\
\bottomrule
\end{tabular}
\end{table}

In most previous references for simplicity, the pump light fields in the FWM process were generally approximated to the invariable constant. When the atomic and environmental conditions of the nonlinear process were defined, the nonlinear coefficient was considered to be constant. However in the real experiments, even the pump light fields are lasers, they obey the Gaussian statistical distributions, the nonlinear coefficient would not be a constant. Thus, when the atomic and environmental conditions of the nonlinear process were defined, even the pump light fields are laser, the statistical type of the generated FWM signal is not exactly the same as its phase conjugation light field, namely probe beam. 
 
As seen in Eq.~(\ref{eq:eight}), we found that when the atomic and environmental conditions of the nonlinear process are defined, the nonlinear coefficients will be dependent on the light fields of pump beams ${E_m}$ and ${E_n}$. The fluctuating nonlinear coefficient of interaction results from the combination of pump light fields. By combining different statistical distributions, the fluctuation degree of the nonlinear coefficient is controllable, then the affected modulation and the increased light intensity fluctuations of the output beam would be tunable flexibly.

Furthermore, there are three same thermal light fields with synchronous fluctuations through FWM in the 3rd-order case. It is interesting to study whether the intensity fluctuations of output beam can be amplified or not when the incident beams are the same light fields without synchronous fluctuations. Therefore, to amplify intensity fluctuations of output beams in the nonlinear process, the importance of the synchronous fluctuation of incident coupling beams is worth studying.

\section{Conclusion}
\label{sec:Conclusion}

Since amplifying light intensity fluctuations plays an important role in many optical applications, we proposed an Nth-order light intensity fluctuation amplifier through N-order optical processes. In the $N$-order nonlinear processes, $k$ ($0 \le k < 2/N$) incident beams are the laser or coherent light, the light intensity fluctuations of all output beams after the nonlinear interaction will be enhanced, that realizes an Nth-order light intensity fluctuation amplifier. In addition, the most effective way to improve the light intensity fluctuations is that N incident beams are all the same light fields with synchronous fluctuations. The FWM process pumped by different statistical distributions was chosen as experimental verification of this amplifier, the ratio $R$ of the statistical light intensity distribution and the degree of second-order coherence ${g^{(2)}}(0)$ of beams are used to evaluate the affected modulation degree and the increased light intensity fluctuation. The experimental results of the FWM process are consistent with the proposed N-order amplifier. 

By inputting light beams configures with different statistical distributions into N-order optical nonlinear processes, the amplification of light intensity fluctuations is caused by not only the fluctuating light fields of incident coupling beams but also the fluctuating nonlinear coefficient of interaction. The different 1~Nth-order cases exhibit different amplification degrees, the amplification of light intensity fluctuations is tunable and multiplex. The proposed Nth-order light intensity fluctuation amplifier could be widely used in many N-order nonlinear optical effects, including optical harmonics, electromagnetically induced transparency, and four-wave mixing.

\section*{Declaration of competing interest}

The authors declare that they have no known competing financial interests or personal relationships that could have appeared to influence the work reported in this paper. 

\section*{Funding}

Shaanxi Key Research and Development Project (Grant No. 2019ZDLGY09-10); Key Innovation Team of Shaanxi Province (Grant No. 2018TD-024); National Natural Science Foundation of China (Grant No. 61901353).

\bibliographystyle{ref}

\bibliography{reference}

\begin{thebibliography}{38}
\expandafter\ifx\csname natexlab\endcsname\relax\def\natexlab#1{#1}\fi
\providecommand{\url}[1]{\texttt{#1}}
\providecommand{\href}[2]{#2}
\providecommand{\path}[1]{#1}
\providecommand{\DOIprefix}{doi:}
\providecommand{\ArXivprefix}{arXiv:}
\providecommand{\URLprefix}{URL: }
\providecommand{\Pubmedprefix}{pmid:}
\providecommand{\doi}[1]{\href{http://dx.doi.org/#1}{\path{#1}}}
\providecommand{\Pubmed}[1]{\href{pmid:#1}{\path{#1}}}
\providecommand{\bibinfo}[2]{#2}
\ifx\xfnm\relax \def\xfnm[#1]{\unskip,\space#1}\fi
\bibitem[{Kolobov(2007)}]{kolobov2007quantum}
\bibinfo{author}{M.~I. Kolobov}, \bibinfo{title}{Quantum imaging},
  \bibinfo{publisher}{Springer Science \& Business Media},
  \bibinfo{year}{2007}.
\bibitem[{Zeng et~al.(2013)Zeng, Wang, Feng, Liang, and Yang}]{zeng2013laser}
\bibinfo{author}{Y.~Zeng}, \bibinfo{author}{M.~Wang},
  \bibinfo{author}{G.~Feng}, \bibinfo{author}{X.~Liang},
  \bibinfo{author}{G.~Yang},
\newblock \bibinfo{title}{Laser speckle imaging based on intensity fluctuation
  modulation},
\newblock \bibinfo{journal}{Optics letters} \bibinfo{volume}{38}
  (\bibinfo{year}{2013}) \bibinfo{pages}{1313--1315}.
\bibitem[{Boyer et~al.(2008)Boyer, Marino, Pooser, and
  Lett}]{boyer2008entangled}
\bibinfo{author}{V.~Boyer}, \bibinfo{author}{A.~M. Marino},
  \bibinfo{author}{R.~C. Pooser}, \bibinfo{author}{P.~D. Lett},
\newblock \bibinfo{title}{Entangled images from four-wave mixing},
\newblock \bibinfo{journal}{Science} \bibinfo{volume}{321}
  (\bibinfo{year}{2008}) \bibinfo{pages}{544--547}.
\bibitem[{Kolobov and Fabre(2000)}]{PhysRevLett.85.3789}
\bibinfo{author}{M.~I. Kolobov}, \bibinfo{author}{C.~Fabre},
\newblock \bibinfo{title}{Quantum limits on optical resolution},
\newblock \bibinfo{journal}{Phys. Rev. Lett.} \bibinfo{volume}{85}
  (\bibinfo{year}{2000}) \bibinfo{pages}{3789--3792}.
\bibitem[{Sprigg et~al.(2016)Sprigg, Peng, and Shih}]{sprigg2016super}
\bibinfo{author}{J.~Sprigg}, \bibinfo{author}{T.~Peng},
  \bibinfo{author}{Y.~Shih},
\newblock \bibinfo{title}{Super-resolution imaging using the spatial-frequency
  filtered intensity fluctuation correlation},
\newblock \bibinfo{journal}{Scientific reports} \bibinfo{volume}{6}
  (\bibinfo{year}{2016}) \bibinfo{pages}{1--7}.
\bibitem[{Sroda et~al.(2020)Sroda, Makowski, Tenne, Rossman, Lubin, Oron, and
  Lapkiewicz}]{sroda2020sofism}
\bibinfo{author}{A.~Sroda}, \bibinfo{author}{A.~Makowski},
  \bibinfo{author}{R.~Tenne}, \bibinfo{author}{U.~Rossman},
  \bibinfo{author}{G.~Lubin}, \bibinfo{author}{D.~Oron},
  \bibinfo{author}{R.~Lapkiewicz},
\newblock \bibinfo{title}{Sofism: Super-resolution optical fluctuation image
  scanning microscopy},
\newblock \bibinfo{journal}{Optica} \bibinfo{volume}{7} (\bibinfo{year}{2020})
  \bibinfo{pages}{1308--1316}.
\bibitem[{Digman et~al.(2005)Digman, Sengupta, Wiseman, Brown, Horwitz, and
  Gratton}]{DIGMAN2005L33}
\bibinfo{author}{M.~A. Digman}, \bibinfo{author}{P.~Sengupta},
  \bibinfo{author}{P.~W. Wiseman}, \bibinfo{author}{C.~M. Brown},
  \bibinfo{author}{A.~R. Horwitz}, \bibinfo{author}{E.~Gratton},
\newblock \bibinfo{title}{Fluctuation correlation spectroscopy with a
  laser-scanning microscope: Exploiting the hidden time structure},
\newblock \bibinfo{journal}{Biophysical Journal} \bibinfo{volume}{88}
  (\bibinfo{year}{2005}) \bibinfo{pages}{L33--L36}.
\bibitem[{Ferri et~al.(2005)Ferri, Magatti, Gatti, Bache, Brambilla, and
  Lugiato}]{ferri2005high}
\bibinfo{author}{F.~Ferri}, \bibinfo{author}{D.~Magatti},
  \bibinfo{author}{A.~Gatti}, \bibinfo{author}{M.~Bache},
  \bibinfo{author}{E.~Brambilla}, \bibinfo{author}{L.~A. Lugiato},
\newblock \bibinfo{title}{High-resolution ghost image and ghost diffraction
  experiments with thermal light},
\newblock \bibinfo{journal}{Physical Review Letters} \bibinfo{volume}{94}
  (\bibinfo{year}{2005}) \bibinfo{pages}{183602}.
\bibitem[{Liu et~al.(2013)Liu, Li, Yao, Yu, Zhai, and Wu}]{liu2013high}
\bibinfo{author}{X.-F. Liu}, \bibinfo{author}{M.-F. Li}, \bibinfo{author}{X.-R.
  Yao}, \bibinfo{author}{W.-K. Yu}, \bibinfo{author}{G.-J. Zhai},
  \bibinfo{author}{L.-A. Wu},
\newblock \bibinfo{title}{High-visibility ghost imaging from artificially
  generated non-gaussian intensity fluctuations},
\newblock \bibinfo{journal}{AIP Advances} \bibinfo{volume}{3}
  (\bibinfo{year}{2013}) \bibinfo{pages}{052121}.
\bibitem[{Bromberg et~al.(2009)Bromberg, Katz, and
  Silberberg}]{bromberg2009ghost}
\bibinfo{author}{Y.~Bromberg}, \bibinfo{author}{O.~Katz},
  \bibinfo{author}{Y.~Silberberg},
\newblock \bibinfo{title}{Ghost imaging with a single detector},
\newblock \bibinfo{journal}{Physical Review A} \bibinfo{volume}{79}
  (\bibinfo{year}{2009}) \bibinfo{pages}{053840}.
\bibitem[{Shapiro(2008)}]{shapiro2008computational}
\bibinfo{author}{J.~H. Shapiro},
\newblock \bibinfo{title}{Computational ghost imaging},
\newblock \bibinfo{journal}{Physical Review A} \bibinfo{volume}{78}
  (\bibinfo{year}{2008}) \bibinfo{pages}{061802}.
\bibitem[{Gatti et~al.(2004)Gatti, Brambilla, Bache, and
  Lugiato}]{gatti2004ghost}
\bibinfo{author}{A.~Gatti}, \bibinfo{author}{E.~Brambilla},
  \bibinfo{author}{M.~Bache}, \bibinfo{author}{L.~A. Lugiato},
\newblock \bibinfo{title}{Ghost imaging with thermal light: comparing
  entanglement and classicalcorrelation},
\newblock \bibinfo{journal}{Physical Review Letters} \bibinfo{volume}{93}
  (\bibinfo{year}{2004}) \bibinfo{pages}{093602}.
\bibitem[{Treps et~al.(2003)Treps, Grosse, Bowen, Fabre, Bachor, and
  Lam}]{treps2003quantum}
\bibinfo{author}{N.~Treps}, \bibinfo{author}{N.~Grosse}, \bibinfo{author}{W.~P.
  Bowen}, \bibinfo{author}{C.~Fabre}, \bibinfo{author}{H.-A. Bachor},
  \bibinfo{author}{P.~K. Lam},
\newblock \bibinfo{title}{A quantum laser pointer},
\newblock \bibinfo{journal}{Science} \bibinfo{volume}{301}
  (\bibinfo{year}{2003}) \bibinfo{pages}{940--943}.
\bibitem[{Glauber(1963)}]{PhysRev.130.2529}
\bibinfo{author}{R.~J. Glauber},
\newblock \bibinfo{title}{The quantum theory of optical coherence},
\newblock \bibinfo{journal}{Phys. Rev.} \bibinfo{volume}{130}
  (\bibinfo{year}{1963}) \bibinfo{pages}{2529--2539}.
\bibitem[{Mandel and Wolf(1995)}]{mandel1995optical}
\bibinfo{author}{L.~Mandel}, \bibinfo{author}{E.~Wolf}, \bibinfo{title}{Optical
  coherence and quantum optics}, \bibinfo{publisher}{Cambridge University},
  \bibinfo{year}{1995}.
\bibitem[{You et~al.(2020)You, Quiroz-Ju{\'a}rez, Lambert, Bhusal, Dong,
  Perez-Leija, Javaid, Le{\'o}n-Montiel, and
  Maga{\~n}a-Loaiza}]{you2020identification}
\bibinfo{author}{C.~You}, \bibinfo{author}{M.~A. Quiroz-Ju{\'a}rez},
  \bibinfo{author}{A.~Lambert}, \bibinfo{author}{N.~Bhusal},
  \bibinfo{author}{C.~Dong}, \bibinfo{author}{A.~Perez-Leija},
  \bibinfo{author}{A.~Javaid}, \bibinfo{author}{R.~d.~J. Le{\'o}n-Montiel},
  \bibinfo{author}{O.~S. Maga{\~n}a-Loaiza},
\newblock \bibinfo{title}{Identification of light sources using machine
  learning},
\newblock \bibinfo{journal}{Applied Physics Reviews} \bibinfo{volume}{7}
  (\bibinfo{year}{2020}) \bibinfo{pages}{021404}.
\bibitem[{Spasibko et~al.(2017)Spasibko, Kopylov, Krutyanskiy, Murzina, Leuchs,
  and Chekhova}]{PhysRevLett.119.223603}
\bibinfo{author}{K.~Y. Spasibko}, \bibinfo{author}{D.~A. Kopylov},
  \bibinfo{author}{V.~L. Krutyanskiy}, \bibinfo{author}{T.~V. Murzina},
  \bibinfo{author}{G.~Leuchs}, \bibinfo{author}{M.~V. Chekhova},
\newblock \bibinfo{title}{Multiphoton effects enhanced due to ultrafast
  photon-number fluctuations},
\newblock \bibinfo{journal}{Phys. Rev. Lett.} \bibinfo{volume}{119}
  (\bibinfo{year}{2017}) \bibinfo{pages}{223603}.
\bibitem[{Akhmediev et~al.(2016)Akhmediev, Kibler, Baronio, Beli{\'c}, Zhong,
  Zhang, Chang, Soto-Crespo, Vouzas, Grelu et~al.}]{akhmediev2016roadmap}
\bibinfo{author}{N.~Akhmediev}, \bibinfo{author}{B.~Kibler},
  \bibinfo{author}{F.~Baronio}, \bibinfo{author}{M.~Beli{\'c}},
  \bibinfo{author}{W.-P. Zhong}, \bibinfo{author}{Y.~Zhang},
  \bibinfo{author}{W.~Chang}, \bibinfo{author}{J.~M. Soto-Crespo},
  \bibinfo{author}{P.~Vouzas}, \bibinfo{author}{P.~Grelu}, et~al.,
\newblock \bibinfo{title}{Roadmap on optical rogue waves and extreme events},
\newblock \bibinfo{journal}{Journal of Optics} \bibinfo{volume}{18}
  (\bibinfo{year}{2016}) \bibinfo{pages}{063001}.
\bibitem[{Zhang et~al.(2019)Zhang, Lu, Zhou, Zhang, Li, and
  Zhang}]{zhang2019superbunching}
\bibinfo{author}{L.~Zhang}, \bibinfo{author}{Y.~Lu}, \bibinfo{author}{D.~Zhou},
  \bibinfo{author}{H.~Zhang}, \bibinfo{author}{L.~Li},
  \bibinfo{author}{G.~Zhang},
\newblock \bibinfo{title}{Superbunching effect of classical light with a
  digitally designed spatially phase-correlated wave front},
\newblock \bibinfo{journal}{Physical Review A} \bibinfo{volume}{99}
  (\bibinfo{year}{2019}) \bibinfo{pages}{063827}.
\bibitem[{Lambropoulos et~al.(1966)Lambropoulos, Kikuchi, and
  Osborn}]{lambropoulos1966coherence}
\bibinfo{author}{P.~Lambropoulos}, \bibinfo{author}{C.~Kikuchi},
  \bibinfo{author}{R.~K. Osborn},
\newblock \bibinfo{title}{Coherence and two-photon absorption},
\newblock \bibinfo{journal}{Physical Review} \bibinfo{volume}{144}
  (\bibinfo{year}{1966}) \bibinfo{pages}{1081}.
\bibitem[{Jechow et~al.(2013)Jechow, Seefeldt, Kurzke, Heuer, and
  Menzel}]{jechow2013enhanced}
\bibinfo{author}{A.~Jechow}, \bibinfo{author}{M.~Seefeldt},
  \bibinfo{author}{H.~Kurzke}, \bibinfo{author}{A.~Heuer},
  \bibinfo{author}{R.~Menzel},
\newblock \bibinfo{title}{Enhanced two-photon excited fluorescence from imaging
  agents using true thermal light},
\newblock \bibinfo{journal}{Nature Photonics} \bibinfo{volume}{7}
  (\bibinfo{year}{2013}) \bibinfo{pages}{973--976}.
\bibitem[{Lecompte et~al.(1975)Lecompte, Mainfray, Manus, and
  Sanchez}]{lecompte1975laser}
\bibinfo{author}{C.~Lecompte}, \bibinfo{author}{G.~Mainfray},
  \bibinfo{author}{C.~Manus}, \bibinfo{author}{F.~Sanchez},
\newblock \bibinfo{title}{Laser temporal-coherence effects on multiphoton
  ionization processes},
\newblock \bibinfo{journal}{Physical Review A} \bibinfo{volume}{11}
  (\bibinfo{year}{1975}) \bibinfo{pages}{1009}.
\bibitem[{Mouloudakis and Lambropoulos(2019)}]{mouloudakis2019revisiting}
\bibinfo{author}{G.~Mouloudakis}, \bibinfo{author}{P.~Lambropoulos},
\newblock \bibinfo{title}{Revisiting photon-statistics effects on multiphoton
  ionization. ii. connection to realistic systems},
\newblock \bibinfo{journal}{Physical Review A} \bibinfo{volume}{99}
  (\bibinfo{year}{2019}) \bibinfo{pages}{063419}.
\bibitem[{Lamprou et~al.(2020)Lamprou, Liontos, Papadakis, and
  Tzallas}]{lamprou2020perspective}
\bibinfo{author}{T.~Lamprou}, \bibinfo{author}{I.~Liontos},
  \bibinfo{author}{N.~Papadakis}, \bibinfo{author}{P.~Tzallas},
\newblock \bibinfo{title}{A perspective on high photon flux nonclassical light
  and applications in nonlinear optics},
\newblock \bibinfo{journal}{High power laser science and engineering}
  \bibinfo{volume}{8} (\bibinfo{year}{2020}).
\bibitem[{Hammani et~al.(2009)Hammani, Finot, and Millot}]{Hammani:09}
\bibinfo{author}{K.~Hammani}, \bibinfo{author}{C.~Finot},
  \bibinfo{author}{G.~Millot},
\newblock \bibinfo{title}{Emergence of extreme events in fiber-based parametric
  processes driven by a partially incoherent pump wave},
\newblock \bibinfo{journal}{Opt. Lett.} \bibinfo{volume}{34}
  (\bibinfo{year}{2009}) \bibinfo{pages}{1138--1140}.
\bibitem[{Zhou et~al.(2017)Zhou, Li, Bai, Chen, Liu, Xu, and
  Zheng}]{zhou2017superbunching}
\bibinfo{author}{Y.~Zhou}, \bibinfo{author}{F.~Li}, \bibinfo{author}{B.~Bai},
  \bibinfo{author}{H.~Chen}, \bibinfo{author}{J.~Liu}, \bibinfo{author}{Z.~Xu},
  \bibinfo{author}{H.~Zheng},
\newblock \bibinfo{title}{Superbunching pseudothermal light},
\newblock \bibinfo{journal}{Physical Review A} \bibinfo{volume}{95}
  (\bibinfo{year}{2017}) \bibinfo{pages}{053809}.
\bibitem[{Straka et~al.(2018)Straka, Mika, and
  Je{\v{z}}ek}]{straka2018generator}
\bibinfo{author}{I.~Straka}, \bibinfo{author}{J.~Mika},
  \bibinfo{author}{M.~Je{\v{z}}ek},
\newblock \bibinfo{title}{Generator of arbitrary classical photon statistics},
\newblock \bibinfo{journal}{Optics Express} \bibinfo{volume}{26}
  (\bibinfo{year}{2018}) \bibinfo{pages}{8998--9010}.
\bibitem[{Zhou et~al.(2019)Zhou, Zhang, Wang, Zhang, Chen, Zheng, Liu, Li, and
  Xu}]{zhou2019superbunching}
\bibinfo{author}{Y.~Zhou}, \bibinfo{author}{X.~Zhang},
  \bibinfo{author}{Z.~Wang}, \bibinfo{author}{F.~Zhang},
  \bibinfo{author}{H.~Chen}, \bibinfo{author}{H.~Zheng},
  \bibinfo{author}{J.~Liu}, \bibinfo{author}{F.~Li}, \bibinfo{author}{Z.~Xu},
\newblock \bibinfo{title}{Superbunching pseudothermal light with intensity
  modulated laser light and rotating groundglass},
\newblock \bibinfo{journal}{Optics Communications} \bibinfo{volume}{437}
  (\bibinfo{year}{2019}) \bibinfo{pages}{330--336}.
\bibitem[{Perina(1993)}]{perina1993photon}
\bibinfo{author}{J.~Perina},
\newblock \bibinfo{title}{Photon statistics of four-wave mixing of nonclassical
  light with pump depletion},
\newblock in: \bibinfo{booktitle}{16th Congress of the International Commission
  for Optics: Optics as a Key to High Technology}, volume
  \bibinfo{volume}{1983}, \bibinfo{organization}{International Society for
  Optics and Photonics}, \bibinfo{year}{1993}, p. \bibinfo{pages}{19830U}.
\bibitem[{Cao et~al.(2016)Cao, Yang, Wang, Qiu, Wei, Gao, and
  Li}]{cao2016resolution}
\bibinfo{author}{M.~Cao}, \bibinfo{author}{X.~Yang}, \bibinfo{author}{J.~Wang},
  \bibinfo{author}{S.~Qiu}, \bibinfo{author}{D.~Wei}, \bibinfo{author}{H.~Gao},
  \bibinfo{author}{F.~Li},
\newblock \bibinfo{title}{Resolution enhancement of ghost imaging in atom
  vapor},
\newblock \bibinfo{journal}{Optics Letters} \bibinfo{volume}{41}
  (\bibinfo{year}{2016}) \bibinfo{pages}{5349--5352}.
\bibitem[{Yu et~al.(2016)Yu, Wang, Liu, Wang, Cao, Wei, Gao, and Li}]{Yu:16}
\bibinfo{author}{Y.~Yu}, \bibinfo{author}{C.~Wang}, \bibinfo{author}{J.~Liu},
  \bibinfo{author}{J.~Wang}, \bibinfo{author}{M.~Cao},
  \bibinfo{author}{D.~Wei}, \bibinfo{author}{H.~Gao}, \bibinfo{author}{F.~Li},
\newblock \bibinfo{title}{Ghost imaging with different frequencies through
  non-degenerated four-wave mixing},
\newblock \bibinfo{journal}{Opt. Express} \bibinfo{volume}{24}
  (\bibinfo{year}{2016}) \bibinfo{pages}{18290--18296}.
\bibitem[{Liu et~al.(2016)Liu, Fang, Zhou, Zhang, Gao, Li, Gao, and
  Li}]{liu2016enhanced}
\bibinfo{author}{R.~Liu}, \bibinfo{author}{A.~Fang}, \bibinfo{author}{Y.~Zhou},
  \bibinfo{author}{P.~Zhang}, \bibinfo{author}{S.~Gao},
  \bibinfo{author}{H.~Li}, \bibinfo{author}{H.~Gao}, \bibinfo{author}{F.~Li},
\newblock \bibinfo{title}{Enhanced visibility of ghost imaging and interference
  using squeezed thermal light},
\newblock \bibinfo{journal}{Physical Review A} \bibinfo{volume}{93}
  (\bibinfo{year}{2016}) \bibinfo{pages}{013822}.
\bibitem[{Manceau et~al.(2019)Manceau, Spasibko, Leuchs, Filip, and
  Chekhova}]{manceau2019indefinite}
\bibinfo{author}{M.~Manceau}, \bibinfo{author}{K.~Y. Spasibko},
  \bibinfo{author}{G.~Leuchs}, \bibinfo{author}{R.~Filip},
  \bibinfo{author}{M.~V. Chekhova},
\newblock \bibinfo{title}{Indefinite-mean pareto photon distribution from
  amplified quantum noise},
\newblock \bibinfo{journal}{Phys. Rev. Lett.} \bibinfo{volume}{123}
  (\bibinfo{year}{2019}) \bibinfo{pages}{123606}.
\bibitem[{Boyd(2003)}]{boyd2003nonlinear}
\bibinfo{author}{R.~W. Boyd}, \bibinfo{title}{Nonlinear optics},
  \bibinfo{publisher}{Elsevier}, \bibinfo{year}{2003}.
\bibitem[{Loudon and Rodney(2000)}]{Loudon1983The}
\bibinfo{author}{Loudon}, \bibinfo{author}{Rodney}, \bibinfo{title}{The Quantum
  Theory of Light}, \bibinfo{publisher}{Oxford University Press},
  \bibinfo{year}{2000}.
\bibitem[{Steck(2001)}]{steck2001rubidium}
\bibinfo{author}{D.~A. Steck}, \bibinfo{title}{Rubidium 87 d line data},
  \bibinfo{year}{2001}.
\bibitem[{Boitier et~al.(2011)Boitier, Godard, Dubreuil, Delaye, Fabre, and
  Rosencher}]{boitier2011photon}
\bibinfo{author}{F.~Boitier}, \bibinfo{author}{A.~Godard},
  \bibinfo{author}{N.~Dubreuil}, \bibinfo{author}{P.~Delaye},
  \bibinfo{author}{C.~Fabre}, \bibinfo{author}{E.~Rosencher},
\newblock \bibinfo{title}{Photon extrabunching in ultrabright twin beams
  measured by two-photon counting in a semiconductor},
\newblock \bibinfo{journal}{Nature Communications} \bibinfo{volume}{2}
  (\bibinfo{year}{2011}) \bibinfo{pages}{1--6}.
\bibitem[{Iskhakov et~al.(2012)Iskhakov, P{\'e}rez, Spasibko, Chekhova, and
  Leuchs}]{iskhakov2012superbunched}
\bibinfo{author}{T.~S. Iskhakov}, \bibinfo{author}{A.~P{\'e}rez},
  \bibinfo{author}{K.~Y. Spasibko}, \bibinfo{author}{M.~Chekhova},
  \bibinfo{author}{G.~Leuchs},
\newblock \bibinfo{title}{Superbunched bright squeezed vacuum state},
\newblock \bibinfo{journal}{Optics Letters} \bibinfo{volume}{37}
  (\bibinfo{year}{2012}) \bibinfo{pages}{1919--1921}.

\end{thebibliography}

\end{document}